\documentclass[twocolumn]{aastex631}
\usepackage{xspace}
\usepackage{pifont}
\usepackage{multirow}

\defcitealias{Poggianti2016JELLYFISHREDSHIFT}{P16}

\received{June 8, 2021}

\shorttitle{The importance of RPS in clusters}
\shortauthors{Vulcani et al.}

\graphicspath{{./}{figures/}}

\begin{document}

\title{The relevance  of ram pressure stripping for the evolution of blue cluster galaxies as seen at optical wavelengths}

\author[0000-0003-0980-1499]{Benedetta Vulcani}
\affiliation{INAF- Osservatorio astronomico di Padova, Vicolo Osservatorio 5, I-35122 Padova, Italy} 

\author[0000-0001-8751-8360]{Bianca M. Poggianti}
\affiliation{INAF- Osservatorio astronomico di Padova, Vicolo Osservatorio 5, I-35122 Padova, Italy}

\author[0000-0001-5303-6830]{Rory Smith}\affiliation{Korea Astronomy and Space Science Institute (KASI), 776 Daedeokdae-ro, Yuseong-gu, Daejeon 34055, Republic of Korea}

\author[0000-0002-1688-482X]{Alessia Moretti}\affiliation{INAF- Osservatorio astronomico di Padova, Vicolo Osservatorio 5, I-35122 Padova, Italy}

\author[0000-0003-2150-1130]{Yara L. Jaff\'e}\affiliation{Instituto de F\'isica y Astronom\'ia, Facultad de Ciencias, Universidad de Valpara\'iso, Avda. Gran Breta\~na 1111, Casilla 5030, Valpara\'iso, Chile}

\author[0000-0002-7296-9780]{Marco Gullieuszik}\affiliation{INAF- Osservatorio astronomico di Padova, Vicolo Osservatorio 5, I-35122 Padova, Italy}

\author[0000-0002-7042-1965]{Jacopo Fritz}
\affiliation{Instituto de Radioastronom\'ia y Astrof\'isica, Universidad Nacional Aut\'onoma de M\'exico, Morelia, Michoac\'an, 58089 M\'exico\\}

\author{Callum Bellhouse}\affiliation{INAF- Osservatorio astronomico di Padova, Vicolo Osservatorio 5, I-35122 Padova, Italy}

\begin{abstract}
Ram pressure stripping is one of the most efficient mechanisms able to affect the gas reservoir in cluster galaxies and in the last decades many studies have characterized the properties of stripped galaxies. A definite census of the importance of this process in local clusters is though still missing. Here we characterize the fraction of galaxies showing signs of stripping at optical wavelengths, using the data of 66 clusters  from the WINGS and OMEGAWINGS surveys. We focus on the infalling galaxy population and hence only consider blue, bright (B$<18.2$) late-type spectroscopically confirmed cluster members within 2 virial radii. In addition to ``traditional'' stripping candidates (SC) -- i.e. galaxies showing unilateral debris and tails -- we also consider unwinding galaxies (UG) as  potentially stripped galaxies. Recent work has indeed unveiled a connection between unwinding features and ram pressure stripping and even though only integral field studies can inform on how often these features are indeed due to ram pressure, it is important to include them in the global census.
We performed a visual inspection of B-band images and here we release a catalog of 143 UG.   
SC and UG each represent $\sim 15-20\%$ of the inspected sample. If we make the assumption that they both are undergoing ram pressure stripping, we can conclude that at any given time in the low-z universe about 35\% of the infalling cluster population show signs of stripping in their morphology at optical wavelengths. These fractions depend on color, mass, morphology, and  little on clustercentric distance. Making some rough assumptions on the duration of the tail visibility and on the time cluster galaxies can maintain blue colors, we infer that almost all bright  blue late-type cluster galaxies undergo a stripping phase during their life, boosting the importance of ram pressure stripping in cluster galaxy evolution. 
\end{abstract}


\section{Introduction} \label{sec:intro}
Ram pressure stripping \citep{Gunn1972} by the Intracluster Medium (ICM) is one of the most advocated mechanisms that affect the properties of cluster galaxies, yet its  overall
efficiency in shaping the cluster galaxy populations has not been definitely quantified yet. Galaxies infalling into clusters  are expected to experience a pressure of ${\displaystyle P_{r}\approx \rho _{e}v^{2}}$
where $P_{r}$ is the ram pressure, $\rho _{e}$ the intracluster gas density, and $v$ the speed of the galaxy relative to the medium.
When the gas in the disk is gravitationally bound to the galaxy less strongly than the force from the ICM `wind' due to the ram pressure, it gets stripped and lost to the ICM. Observationally, ram pressure stripping results in many  effects including compression of gas along the leading edge of the disc \citep[e.g.][]{Rasmussen2006, Poggianti2019b}, disturbed galaxy morphologies and trailing tails of stripped gas \citep[e.g.,][]{vanGorkom2004, Kenney2004, Poggianti2017, Fumagalli2014}, condensation of star forming knots in the tails \citep[e.g.,][]{Kenney2014, Poggianti2019}. Eventually, it can also remove a galaxy’s entire gas supply, making it an important quenching pathway for satellite galaxies \citep{Vollmer2001, Tonnesen2007, Vulcani2020}. 

The most extreme examples of galaxies undergoing ram pressure stripping have been dubbed “jellyfish” galaxies, due to the evocative morphologies of their star forming tails \citep{Smith2010}. 

Recently, an additional aftermath of the ram pressure stripping has been proposed: the galaxy spiral arms can be ``unwound'' or broadened as an effect of the gas stripping by the ICM \citep{Bellhouse2021}. The unwound component is characterised only by younger stars, while older stars in the disc remain undisturbed \citep{Bellhouse2021}. Hydrodynamical simulations support the appearance of unwinding: \cite{SchulzStruck2001} report the formation of flocculent spiral arms in the non-stripped disc gas as the remnant disc is compressed and a  stretching and shearing of outer spiral arms prior to their removal from the galaxy \citep[see also][]{Roediger2014, Steinhauser2016}. 

The origin of the unwinding effect in cluster galaxies in general is however still unsettled. Indeed, the observed unwinding structures resemble the tidal tails due to gravitational interactions that typically occur in the field and dedicated spatially resolved studies are needed to distinguish between the two and quantify how often these features are a signature of ram pressure stripping in clusters. 

In the last decades, much attention has been dedicated to ram pressure stripped galaxies and currently few hundreds galaxies have been confirmed to be undergoing a ram pressure event. A number of studies have characterized in detail individual galaxies, paving the way for further investigation, but until recently they mostly focused on small and sparse samples that prevent a detailed characterization of the incidence of such events \citep[e.g.,][just to cite a few]{Fumagalli2014, Merluzzi2016, Fossati2016, Boselli2016, Consolandi2017, Boselli2018, Owers2012, Boselli2021}. 

From a theoretical point of view, \cite{Yun2019} have used IllustrisTNG simulations to study the demographics and properties of jellyfish
galaxies. They selected jellyfish candidates by visually identifying satellites orbiting in massive groups and clusters exhibiting highly asymmetric distributions of gas and gas tails. Considering only galaxies with $\log M_\ast/M_\sun>9.5$ and within the virial radius, they found that at $z<0.6$ about 13\% of cluster satellites present signatures of ram-pressure stripping and gaseous tails stemming from their main luminous bodies. This number increases to 31\% when only galaxies with gas are considered. They also show how the given value is a lower limit: the random orientation entails a loss of about 30\% of galaxies that in an optimal projection would otherwise be identified as jellyfish. 

From the observational point of view, the two first systematic searches for ram pressure stripping candidates in a large sample of clusters are \cite{McPartland2016} and \citet[from now on P16]
{Poggianti2016JELLYFISHREDSHIFT}. In both cases, optical images were analyzed to identify galaxies whose  morphology suggests they might be experiencing stripping of their gaseous material. \cite{McPartland2016} identified 223  0.3$<z<$0.6 cluster galaxies using quantitative morphological parameters, while \citetalias{Poggianti2016JELLYFISHREDSHIFT} visually inspected optical images of cluster and field galaxies at 0.04$<z<$0.07, assembling a  sample of 419 candidates, 344 of which in clusters. These galaxies are found in all the inspected clusters and at all clustercentric radii, they are disky, have stellar masses ranging from $\log M_\ast/M_\sun \sim 9-11.5$ and the majority of them form stars at a rate that is on average a factor of 2 higher compared to non-stripped galaxies of similar mass. 
The number of  candidates  correlate with neither the cluster velocity dispersion $\sigma$ nor X-ray luminosity, but  \citetalias{Poggianti2016JELLYFISHREDSHIFT}  have not quantified in detail their incidence over the total cluster population. 

Subsequently, other groups have provided quite large catalogs of ram pressure stripped candidates, based on different wavelengths and identification techniques \citep[e.g.,][]{RobertsParker2020, Roberts2021a, Roberts2021b, Durret2021}. \cite{Roberts2021b, Roberts2022} showed how ram pressure stripped galaxies with long radio continuum tails are most commonly found in clusters, with the frequency decreasing towards the lowest-mass groups. 

However, typically  searches in both observations and simulations  select candidates against unwinding features, as they are often not considered as a proxy for ram pressure stripping.\footnote{In \citetalias{Poggianti2016JELLYFISHREDSHIFT}, the few unwinding galaxies entering the sample show additional clear signs of ram pressure stripping, such as extended tails departing from the disc.} They  therefore miss a subpopulation that might play an important role in understanding the efficiency of ram pressure stripping. 
Optical spectroscopic follow ups are  crucial to confirm that ram pressure stripping is the main mechanism producing gas tails and unwinding features and the GAs Stripping Phenomena in Galaxies (GASP, \citealt{Poggianti2017}) program, based on VLT/MUSE observations,  has allowed us to make a significant step forward in our understanding of the peculiar signatures ram pressure  leaves on the spatially resolved galaxy properties. GASP observed 94 (64 cluster and 30 field galaxies) of the 419 ram pressure stripping candidates identified by \citetalias{Poggianti2016JELLYFISHREDSHIFT}, unveiling for the first time the link between ram pressure and unwinding features \citep{Bellhouse2017, Bellhouse2019, Bellhouse2021}.\footnote{In contrast, in the field sample a wide variety of mechanisms can produce similar signatures on the galaxy optical appearance \citep{Vulcani2021}.}

While definitely allowing a change of perspective, the cluster sample observed by GASP is still too small to perform a robust statistical analysis on the incidence of ram pressure stripping among galaxies inhabiting the most massive regions of the universe. 
However, GASP has allowed us to test the reliability of the \citetalias{Poggianti2016JELLYFISHREDSHIFT} classification, reassuring us about the possibility of using the ram pressure stripping candidate parent catalog to estimate the frequency of ram pressure stripping events in clusters (B.M. Poggianti et al. in prep.). 

The main goal of this paper is to quantify the fraction of blue late-type galaxies showing ram pressure stripping signatures at optical wavelengths in local clusters and how this fraction depends on galaxy and cluster properties. To assemble a comprehensive sample of possible stripping candidates, we complement the \citetalias{Poggianti2016JELLYFISHREDSHIFT} catalog with a catalog of unwinding galaxies, presented for the first time here. This choice is dictated by the recent GASP results already mentioned above \citep{Bellhouse2019, Bellhouse2021} that showed that ram pressure stripping can also unwind the spiral arms of cluster galaxies.
Only additional studies will establish the real fraction of unwinding galaxies due to ram pressure stripping  rather than tidal interactions; therefore including them in the ram pressure census provides an upper limit of the real fraction. 

We note that the results presented here are based on the information coming from the optical images, therefore strongly rely on a specific class of ram pressure stripped galaxies, such as galaxies with tails bright in the visible, suggesting rather recent extraplanar star formation. Different results might be obtained relying on different set of images and wavelengths, such as atomic hydrogen -- typically even more sensitive to stripping, but not directly related to star formation -- or UV light -- more sensitive to star formation taking place during the last few hundreds of Myr -- or radio continuum -- due to the presence of magnetic fields whose origin and properties are still poorly constrained. Currently, there are no homogeneous and large enough datasets that allow to estimate the incidence of ram pressure stripped galaxies combining the information coming from complementary multi-wavelength observations. 

Throughout the paper, we adopt a \cite{Chabrier2003} initial mass function (IMF) in the mass range 0.1-100 M$_{\odot}$. The cosmological constants assumed are $\Omega_m=0.3$, $\Omega_{\Lambda}=0.7$ and H$_0=70$ km s$^{-1}$ Mpc$^{-1}$. 

\section{Dataset} \label{sec:data}
All galaxies considered in this paper have been observed in the context of the WINGS and OMEGAWINGS surveys, which target clusters of galaxies selected on the basis of their X-ray luminosity \citep{ebeling96, ebeling98, ebeling00}, covering a wide range in cluster masses ($500<\sigma [km/s]<1200$, $43.3<\log(L_X  [erg/s])<45$, \citealt{Fasano2006}), in the redshift range $0.04<z<0.07$. 

The WINGS dataset \citep{Fasano2006, Moretti2014WINGSClusters}  consists of B and V deep photometry of a $34^\prime \times 34^\prime$ field of view with the WFC@INT and the WFC@2.2mMPG/ESO \citep{varela09} for 76 clusters, along with J and K imaging with WFC@UKIRT \citep{Valentinuzzi2009} and some U-band imaging \citep{omizzolo13, Donofrio2020}.  OMEGAWINGS extends WINGS observations,  quadrupling the area covered (1 square degree) for 46 of the 76 clusters, allowing to reach up to $\sim$2.5 cluster virial radii. It is based on  B and V observations with OmegaCAM@VST GTO programs \citep{Gullieuszik2015}. OMEGAWINGS photometry is in general 0.5–1.0 mag shallower than that of WINGS, but  OMEGAWINGS observations were carried out with a better seeing, so the depth of observations are similar. The OMEGAWINGS magnitude distribution peaks at V$\sim$22.5 mag and that of WINGS at V$\sim$ 23.4 \citep{Gullieuszik2015}. 

V-band images are used to compute morphological types  using MORPHOT, an automatic tool for galaxy morphology, purposely devised in the framework of the WINGS project \citep{Fasano2012}. This catalog is complete for 
galaxies which cover more than 200 pixels in the image.

For both surveys, a number of spectroscopic follow-ups have been conducted, using 2dF@AAT and WYFFOS@WHT observations for WINGS \citep{Cava2009}, and  AAOmega@AAT observations for OMEGAWINGS \citep{Moretti2017}. 
Overall, these spectroscopic follow ups reach very high spectroscopic completeness levels for galaxies brighter than V=20 from the cluster cores to their periphery \citep{Moretti2017}, allowing to robustly determine cluster properties. Cluster redshifts $z_{cl}$ and velocity dispersions $\sigma$ are presented in  \cite{Biviano2017} and \cite{Moretti2017}; the virial radii $R_{200}$ are presented in  \cite{Biviano2017}  and \cite{Gullieuszik2020}. 

To further increase the spectroscopic sample, we carried out a systematic literature search of the available redshifts for the galaxies in WINGS+OMEGAWINGS photometric sample. The search includes data from SIMBAD\footnote{\url{http://simbad.u-strasbg.fr/simbad/}}, the SDSS-DR12\footnote{\url{https://www.sdss.org/dr12/}}, NED\footnote{\url{https://ned.ipac.caltech.edu}}, the Shapley Supercluster Survey \citep[SHASS,][]{Merluzzi2015} and from the National Optical Astronomy Observatory (NOAO)\footnote{\url{https://www.noao.edu}}. For Abell2626 we also consider the recently published redshift catalog by \cite{Healy2021}. The final spectroscopic sample includes 46700 redshifts. 

We compute cluster membership for all the galaxies with redshift, assuming they are members if they are within 3$\sigma$ from the cluster redshift.\footnote{Note that this cut is rather conservative, as beautiful ram pressure stripped galaxies, such as JO201,  are moving with even higher speed \citep{Bellhouse2017}.} 

Total and aperture rest frame magnitudes and colors are computed for all galaxies of the photometric sample assuming they are at the  redshift of the cluster they belong to, using the total (SE{\footnotesize XTRACTOR} AUTO) and B and V magnitudes and the aperture B and V magnitudes measured within a diameter of 10 kpc around each galaxy light barycenter, respectively \citep{Moretti2014WINGSClusters, Gullieuszik2015}. All values are corrected for distance modulus and foreground Galaxy extinction, and k-corrected using tabulated values from \cite{Poggianti97}. 

Stellar masses are derived as in \cite{Vulcani2011}, following \cite{BellDeJong2001} and exploiting the correlation between stellar mass-to-light (M/L) ratio and optical colours of the integrated stellar populations. The total luminosity, $L_{B}$, is derived from the total rest frame  B magnitude; the rest-frame $(B-V)$ color from the aperture rest frame magnitudes.  Then, we use the  equation $\log(M/L_B)=a_B +b_B\times (B-V)$, for the Bruzual \& Charlot model with a \citet{salpeter55} IMF (0.1-125 M$_\odot$) and solar metallicity, $a_B$ = -0.51 and $b_B$ = 1.45. Finally, we scale our masses to a \citet{Chabrier2003} IMF adding -0.24 dex to the logarithmic value of the masses. 

\begin{figure}
    \centering
    \includegraphics[scale=0.6]{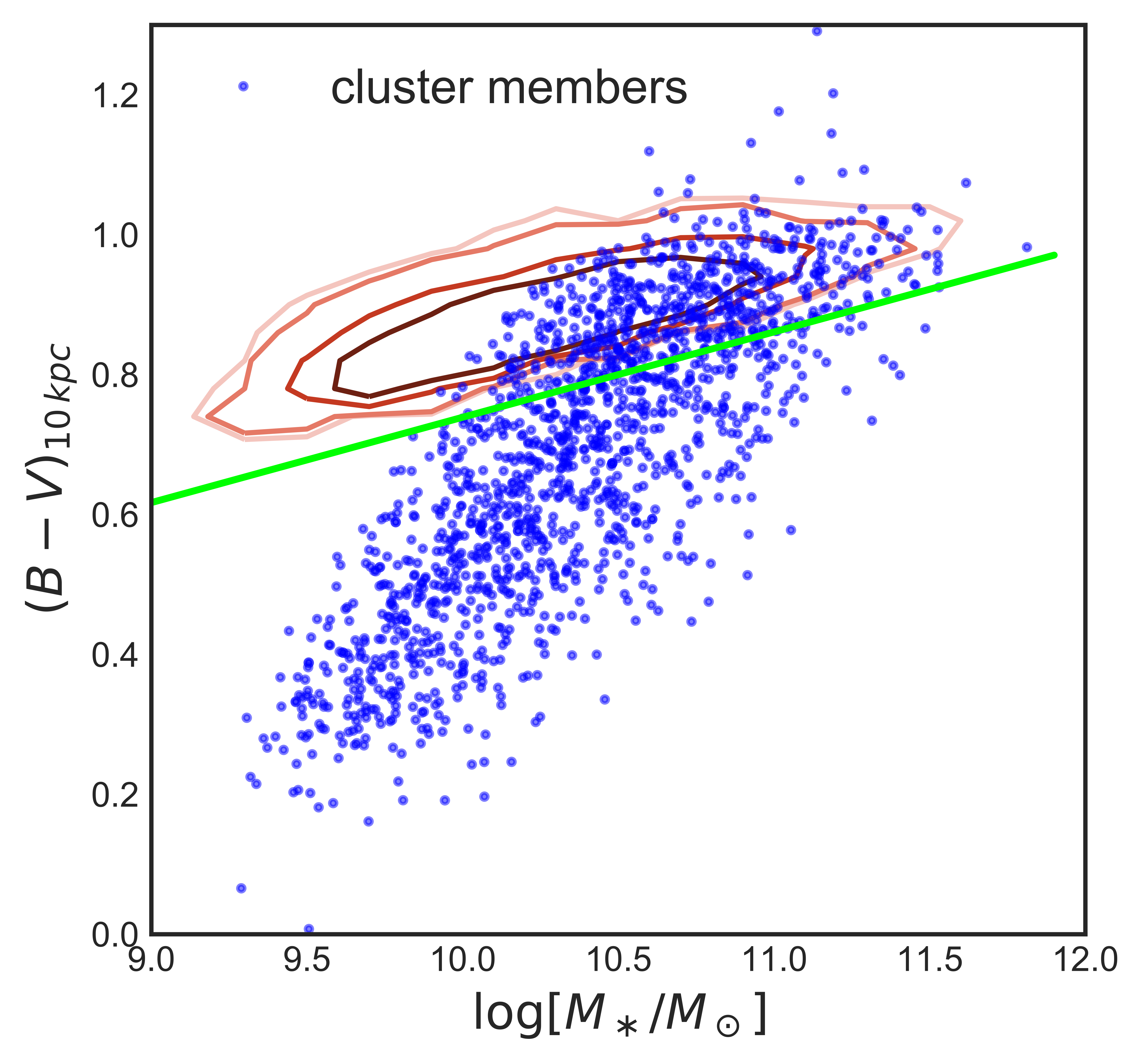}
    \caption{Rest-frame (B-V) - mass diagram for B$<18.2$ late-type cluster members. The red lines represent the red sequence as density plot (see text for details). The green line represents the threshold adopted to separate blue from red galaxies. 
    \label{fig:cmd}}
\end{figure}

\section{Galaxy samples} \label{sec:samples}
Our goal is to quantify the incidence of galaxies showing ram pressure stripping features  in B-band images of local clusters and the natural starting point is the stripping candidate sample presented in \citetalias{Poggianti2016JELLYFISHREDSHIFT}. This sample includes galaxies identified with visual inspection of WINGS and OMEGAWINGS B-band images that are being stripped of their gas and show varying degrees of
morphological evidence for stripping. The catalog does not include galaxies with morphologies clearly disturbed due to mergers and tidal interactions. 
\citetalias{Poggianti2016JELLYFISHREDSHIFT} assigned candidates to five classes according to the visual evidence for stripping signatures in the optical bands (JClass), from extreme cases (JClass 5) to progressively weaker cases, down to the weakest (JClass 1).

\begin{table*}
\caption{Newly identified stripping candidates. Name: progressive identification number; Cluster: cluster the galaxy is member of, (RA, DEC): J2000 coordinates of the galaxy, z: galaxy redshift; JClass: Stripping class a assigned to the galaxy, following the scheme presented in \citetalias{Poggianti2016JELLYFISHREDSHIFT}; $F_d$:  disturbance flag (see text for details); Image: Inspected image: O=OMEGAWINGS (OmegaCAM) or W=WINGS (INT or 2.2m); ID\_WINGS: id in the WINGS/OMEGAWINGS catalog. \label{tab:SC}}
\begin{center}
\begin{tabular}{llrrrrrll}
\hline
\hline
  \multicolumn{1}{c}{Name} &
  \multicolumn{1}{c}{Cluster} &
  \multicolumn{1}{c}{RA[J2000]} &
  \multicolumn{1}{c}{DEC[J2000]} &
  \multicolumn{1}{c}{z} &
  \multicolumn{1}{c}{JCLASS} &
  \multicolumn{1}{c}{$F_{d}$} &
  \multicolumn{1}{c}{Image} &
  \multicolumn{1}{c}{ID\_WINGS} \\
\hline
  SC1 & A1069 & 160.08182 & -8.40895 & 0.0670 & 1 & 1 & OW & WINGSJ104019.62-082431.6\\
  SC2 & A151 & 17.19633 & -15.10619 & 0.0495 & 2 & 1 & OW & WINGSJ010847.12-150622.3\\
  SC3 & A168 & 18.81805 & 0.23531 & 0.0435 & 1 &  & OW & WINGSJ011516.36+001407.6\\
  SC4 & A193 & 21.47709 & 9.12167 & 0.0497 & 1 & 1 & OW & WINGSJ012554.50+090718.0\\
  SC5 & A1983 & 222.82333 & 16.47469 & 0.0395 & 1 &  & OW & WINGSJ145117.61+162829.2\\
\hline
\end{tabular}
\end{center}
\tablecomments{This table is published in its entirety in machine-readable format.  A portion is shown here for guidance regarding its form and content.}
\end{table*}

Nonetheless, as mentioned above, this sample is -by construction- biased against galaxies showing unwinding features, as at the time the catalog was published unwinding features were not recognized as a  possible by-product of ram pressure stripping. Therefore, to obtain a comprehensive census of galaxies  possibly undergoing  stripping we first conducted a systematic search of galaxies showing unwinding features in the same clusters. First, for our purposes, we consider only galaxies with a robustly measured redshift that are cluster members, within 2$\times R_{200}$,\footnote{All the results of the paper will be the same also adopting a more stringent cut in clustercentric distance, such as 0.7$R_{200}$, which is the typical coverage of the WINGS clusters.} with a SE{\footnotesize XTRACTOR} B AUTO $<18.2$ and having a spiral morphology according to MORPHOT. 
We also apply a cut in color, excluding galaxies redder than 
$(B-V)_{rest-frame} =  0.124\times \log M_\ast/M_\sun -0.468$ (see Fig.\ref{fig:cmd}). 
This threshold is obtained determining the  color–mass red sequence of early type galaxies in WINGS and OMEGAWINGS clusters above V=20. We then fix the intercept 1$\sigma$ below the red sequence. These criteria were obtained inspecting the properties of the \citetalias{Poggianti2016JELLYFISHREDSHIFT} sample, in order to assemble homogeneous samples. The need of selecting only blue late-type galaxies comes from the fact that this population represents the gas rich infalling population which might be the most  susceptible to ram pressure. 
Then, we consider the 71 clusters inspected by  \citetalias{Poggianti2016JELLYFISHREDSHIFT} (see their Table 1), which are characterized by good  B-band imaging (seeing $\leq$ 1.2 arcsec) and extract the 66 clusters with spectroscopic coverage. 987 galaxies constitute our starting sample, 
110 of which belong to the \citetalias{Poggianti2016JELLYFISHREDSHIFT} atlas.
 
Two of us (BV and BMP)  visually inspected  the remaining 877 galaxies with three goals: 1) identify any additional stripping candidates missed by the previous visual inspection; 2) select galaxies with clear unwinding features; 3) assemble a reference sample of blue late-type cluster galaxies not showing any clear signs of ram pressure stripping nor of other mechanisms. The images were first inspected independently by each classifier who assigned a JClass following the scheme given in  \citetalias{Poggianti2016JELLYFISHREDSHIFT} to the new stripping candidates and a  UClass to the galaxies with unwinding features (see sec.\ref{sec:un}).  The final classification was agreed upon and a consensus was found on the classification of those galaxies whose individual class differed. 

The classifiers also assigned to all galaxies a disturbance flag $F_d$, to indicate if their morphology appear disturbed in ways that are not suggestive of any specific process.
In the following analysis, we will show some examples and discuss the impact of this population on the final results. 

The classifiers also flagged (66) objects at the borders of the photometric images or very close to bright stars. Indeed, the spatial location of these galaxies would prevent the identification of ram pressure stripped candidates (as done in  \citetalias{Poggianti2016JELLYFISHREDSHIFT}). Following the approach adopted by  \citetalias{Poggianti2016JELLYFISHREDSHIFT},  they also flagged clear cases of post-mergers (33) and mergers/strong interactions (121), as these would hinder the unequivocal identification of ram pressure stripping effects, and galaxies with clear companions producing tidal features (23). All these galaxies (for a total of 240) will be excluded from the following analysis, leaving us with a total sample of 747 galaxies.

\subsection{Stripping candidates (SC)}
We identified additional 35 ram pressure stripping candidates (SC) that were missed by the visual inspection performed by \citetalias{Poggianti2016JELLYFISHREDSHIFT}. Note that one of the three classifiers  of \citetalias{Poggianti2016JELLYFISHREDSHIFT} (BMP) re-inspected the same images, identifying new candidates that were overlooked in the previous inspection. In Tab.\ref{tab:SC} we  provide the list of these new candidates, with a stripping classification coherent with that of \citetalias{Poggianti2016JELLYFISHREDSHIFT}. 
Overall, we detected 1 JClass 4, 1 JClass 3, 12 JClass 2 and 21 JClass 1 galaxies. 

The final sample of stripping candidates considered in this paper is 145, 40 of which are flagged as potentially disturbed by other mechanisms. 

Although only IFU data can indisputably confirm the ram pressure origin of the observed features and thus we cannot exclude the presence of some contaminants -- i.e. galaxies whose extraplanar debris is produced by other mechanisms -- we expect this contamination to be small in clusters based on the GASP results ($< 20$\%, B.M. Poggianti et al. in prep.). 

\subsection{Unwinding galaxies (UG)}\label{sec:un}
\begin{figure*}
    \centering
    \includegraphics[scale=0.45]{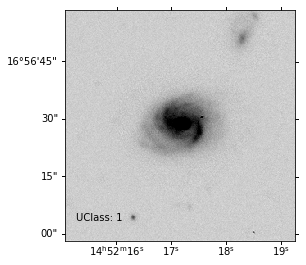}
    \includegraphics[scale=0.45]{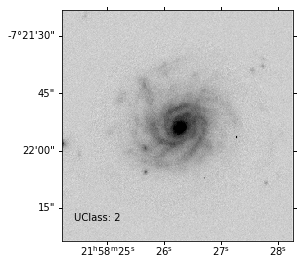}
    \includegraphics[scale=0.45]{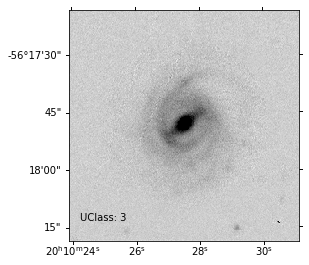}
    \includegraphics[scale=0.45]{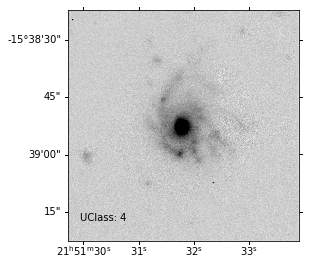}
    \includegraphics[scale=0.45]{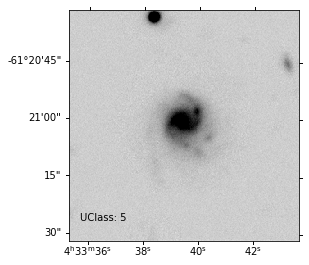}
    \caption{Examples of unwinding galaxies of different classes. The UClass has been assigned to galaxies according to the visual evidence of unwinding features in the optical band, from extreme cases (UClass 5) to progressively weaker cases, down to UClass 1. 
    \label{fig:examples}}
\end{figure*}

We identified 143 galaxies with clear unwinding features (UG). Figure \ref{fig:examples} shows a few examples of these galaxies.
Since the appearance of the pdf figures strongly depends on the screen or printer used and this can make it hard to visualize the features of interest, we  provide cutouts images of each candidate in fits format to allow the reader to display each image with the most appropriate cuts for their screen/printer.
Similarly to the JClass, we tentatively assigned our candidates to five classes according to the visual evidence of unwinding features in the optical band (UClass), from extreme cases (UClass 5) to progressively weaker cases, down to UClass 1. More specifically, these classes carry the indication on how much the arms are ``unrolled'': in UClass 1 galaxies there is a first indication that the arms are opening up, and this opening increases with increasing UClass. For UClass$\geq$3, the unwinding became asymmetric and in UClass 5 cases, at least one long stretched arm is visible on one side. 
As discussed in \citetalias{Poggianti2016JELLYFISHREDSHIFT}, optical signatures are just the tip of the iceberg of distortions in the ionized gas \cite[e.g.,][]{Merluzzi2013, Fumagalli2014}. Therefore, we include in our catalog galaxies over the whole range of degree of evidence for unwinding.

\begin{table*}
\caption{Galaxies showing unwinding features. Name: progressive identification number; Cluster: cluster the galaxy is member of, (RA, DEC): J2000 coordinates of the galaxy, z: galaxy redshift; UClass: Unwinding class a assigned to the galaxy; $F_d$:  disturbance flag (see text for details); Image: Inspected image: O=OMEGAWINGS (OmegaCAM) or W=WINGS (INT or 2.2m); ID\_WINGS: id in the WINGS/OMEGAWINGS catalog. \label{tab:unwind}}
\begin{center}
\begin{tabular}{llrrrrrll}
\hline
\hline
  \multicolumn{1}{c}{Name} &
  \multicolumn{1}{c}{Cluster} &
  \multicolumn{1}{c}{RA[J2000]} &
  \multicolumn{1}{c}{DEC[J2000]} &
  \multicolumn{1}{c}{z} &
  \multicolumn{1}{c}{UCLASS} &
  \multicolumn{1}{c}{$F_{d}$} &
  \multicolumn{1}{c}{Image} &
  \multicolumn{1}{c}{ID\_WINGS} \\
\hline
  UG1 & A1069 & 159.92726 & -8.54181 & 0.0682 & 3 &  & W & WINGSJ103942.54-083230.5\\
  UG2 & A119 & 14.23723 & -1.21173 & 0.0499 & 1 &  & OW & WINGSJ005656.92-011242.5\\
  UG3 & A1291 & 172.68805 & 55.94786 & 0.0509 & 2 & 1 & W & WINGSJ113045.15+555652.8\\
  UG4 & A1291 & 172.83122 & 56.19606 & 0.0526 & 2 & 1 & W & WINGSJ113119.47+561146.1\\
  UG5 & A147 & 17.147297 & 1.943378 & 0.04513 & 2 &  & OW & WINGSJ010835.31+015637.0\\
\hline\end{tabular}
\end{center}
\tablecomments{This table is published in its entirety in machine-readable format.  A portion is shown here for guidance regarding its form and content.}
\end{table*}

Table \ref{tab:unwind} presents the galaxies with unwinding arms. Galaxies are distributed in the Uclasses as follows: 2 UClass 5, 7 UClass 4, 26 Uclass 3, 56 Uclass 2 and 52 UClass 1. Of these, 22 have the disturbance flag set to 1. 

While potentially the different classes could be indicative of a different stage of galaxy evolution, in what follows we will not contrast results for galaxies of different UClasses, both due to the uncertainties of the classification and due to the number of galaxies in each class, especially the highest ones, is very limited, preventing us from obtaining solid conclusions. 

We stress again that this sample is  complementary to the SC sample, as these galaxies do not show clear tails as the galaxies selected by \citetalias{Poggianti2016JELLYFISHREDSHIFT}, but instead show clear signs of unwinding arms. With the current data at our disposal, we can not establish how often the observed features are  due to ram pressure stripping or if these galaxies show tails at other wavelengths. At this stage, the best we can do is to quantify how much this population will impact the final statistics.

\begin{figure*}
    \centering
    \includegraphics[scale=0.45]{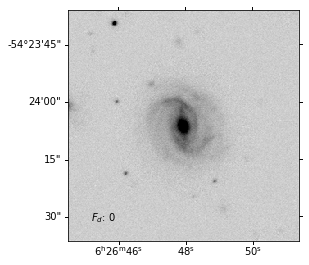}
    \includegraphics[scale=0.45]{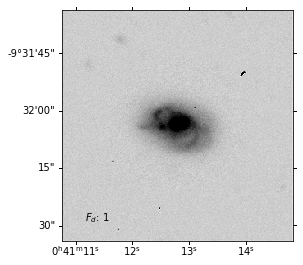}
     \caption{Examples of two reference sample galaxies of different classes. The galaxy on the left has a disturbance flag $F_d$=0, indicating that the galaxy is most likely truly undisturbed, the galaxy on the right has $F_d$=1, indicating that some asymmetries are evident, even though these seem not to be compatible with the typical signatures of ram pressure stripping nor clear tidal interactions or mergers. 
    \label{fig:examples_RS}}
    \end{figure*}

\subsection{The reference sample (RS)}
The reference sample (RS) includes cluster members with no clear signs of stripping nor unwinding arms, for a total of 459 galaxies. 252 of these though are potentially disturbed by other -- not identified -- mechanisms, indicating how rare is to find really undisturbed galaxies in local clusters.  We remind the reader we have on purpose excluded strong interactions, mergers and galaxies with companions (for brevity, ``merging galaxies'' hereafter), for a total of 177 galaxies. In the following we will comment how fraction would change if we included also these galaxies in the sample. Figure \ref{fig:examples_RS} shows two examples of RS galaxies, one with $F_d$=0 and one with $F_d$=1. In the former case, the galaxy is most likely truly undisturbed, in the second case some asymmetries are evident. Nonetheless, these asymmetries seem not to be compatible with the typical signatures of ram pressure stripping nor clear tidal interactions or mergers. As in the case of SC and UG, only IFU data could give more clear indication of the acting mechanisms. 

We stress that this sample is representative of blue late-type cluster galaxies. Field galaxies at similar redshifts and selected following the same criteria could have different properties and be present in a different number, but our scope here is to focus only on the cluster population. Further, as discussed in Sect.\ref{sec:disc}, this sample includes both ``normal'' blue galaxies, recently infallen from the field and currently forming stars undisturbed, and galaxies that are already passive, but still maintain their color for a short time. 

\begin{figure*}
    \centering
    \includegraphics[scale=0.45]{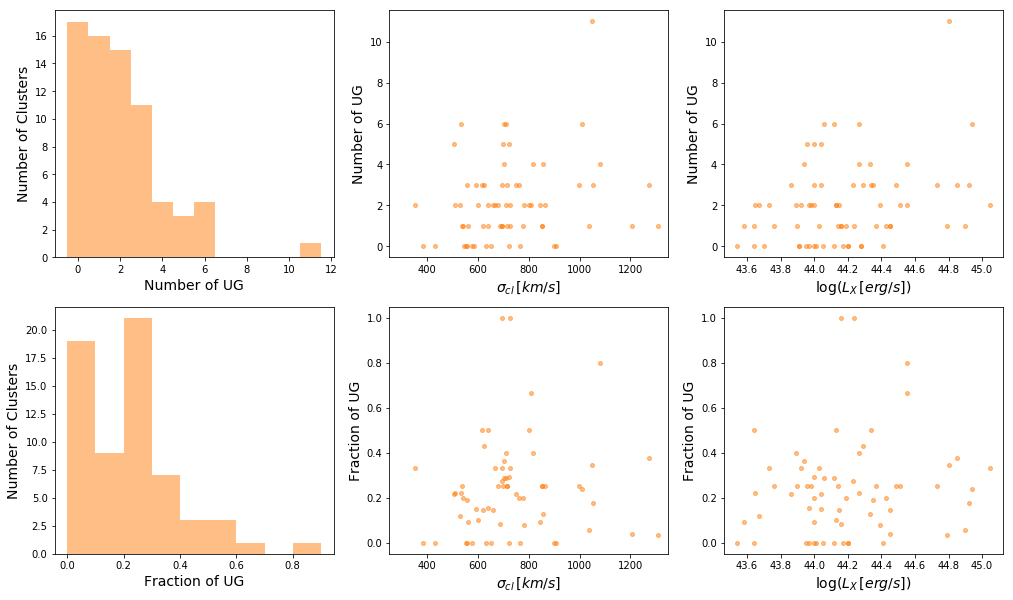}
     \caption{Number (up) and fraction (low) of unwinding galaxies per cluster (left), as a function of the cluster velocity dispersion (middle) and X-ray luminosity (left).
    \label{fig:location_JU}}
\end{figure*}

\section{A view on the unwinding galaxies} \label{sec:properties}

\subsection{The location within the clusters} \label{sec:properties}

\begin{figure*}
    \centering
    \includegraphics[scale=0.53]{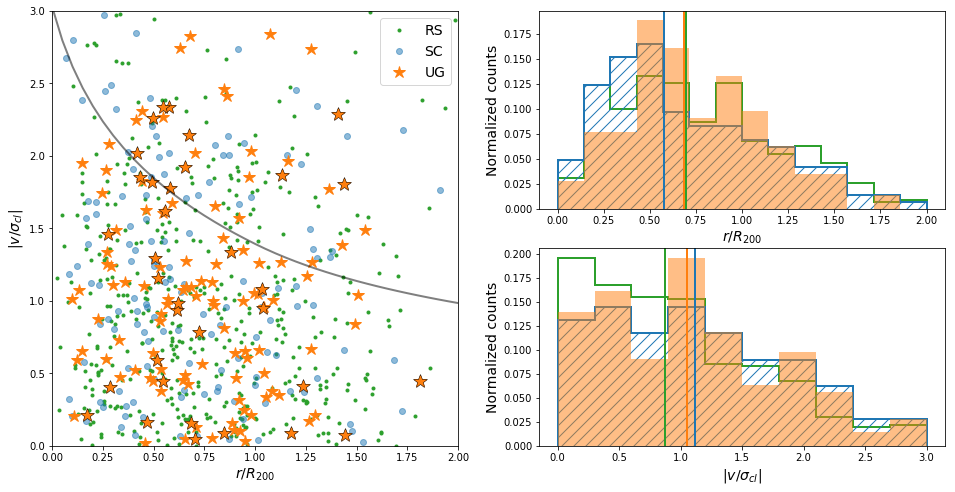}
    \caption{Left: Projected-phase space diagram; right top: distribution of clustercentric distance in unit of $R_{200}$  right bottom: velocity relative to the cluster velocity. Light blue lines and points: stripping candidates, orange lines and stars: unwinding galaxies, green  lines and dots: reference sample. In the left panel, stars surrounded by black line indicate UG with Uclass 3,4,5. The grey curve corresponds to the 3D (un-projected) escape velocity in an NFW halo with concentration c = 6 for reference. This value is appropriate for massive systems \citep{Biviano2017}. In the left panels, vertical lines represent median values.
    \label{fig:pps}}
\end{figure*}

\citetalias{Poggianti2016JELLYFISHREDSHIFT} characterized the location of the SC within the clusters. Here we do the same for the UG. The first panel in Fig. \ref{fig:location_JU} shows that UG are distributed in most of the inspected clusters, even though 16 clusters host no UG. One cluster hosts 11 of them, but on average, each cluster hosts only 2 UG. This number does not correlate with the number of SC (plot not shown), but obviously depends on the number of members per cluster, so the first panel in the second row in Fig. \ref{fig:location_JU} gives the fraction of UG with respect to the total number of blue late-type members in a given cluster (SC+UG+RS). On average, about 20\% of non-interacting cluster blue late-type galaxies are UG.

Next, since our clusters cover a wide range of $\sigma_{cl}$ and $L_X$, we investigate if the number of UG per cluster or their fraction depend on these observables. The second and third panels in both rows of Fig. \ref{fig:location_JU} show that this is not the case. UG are present both in low- and high-mass clusters and no obvious trends are recovered. This result does not change if we only consider the number of UG with a higher class (UClass 3,4,5) or only galaxies with no additional disturbance ($F_d$=0).

We also compare the distribution of  $\sigma_{cl}$ and $L_X$ for the UG, SC, and RS (plots not shown), finding no significant differences. 

UG are observed at all clustercentric radii and span a wide range of relative velocities: Fig.\ref{fig:pps} compares the projected phase space diagram for UG, SC and RS. The distribution of UG is skewed towards larger radii than SC, as shown by the median values in the plot, suggesting that they have fallen more recently  or are on different orbits than galaxies already presenting clear signs of stripping. A Kolmogorov-Smirnov (KS) test probes with high significance that these distributions are different (p$_{value}=0.03$). In contrast, the KS test does not recover any difference between RS and UG and SC, separately. The velocity distributions of the UG and SC samples are quite similar, and the KS test does not find a significant difference. In contrast, the KS test recovers differences between the velocity of RS  and both UG and SC (p$_{value}=0.017$ in both cases). In the next section we will investigate the radial distribution for galaxies removing the effect of stellar mass.

\begin{figure*}
    \centering
    \includegraphics[scale=0.35]{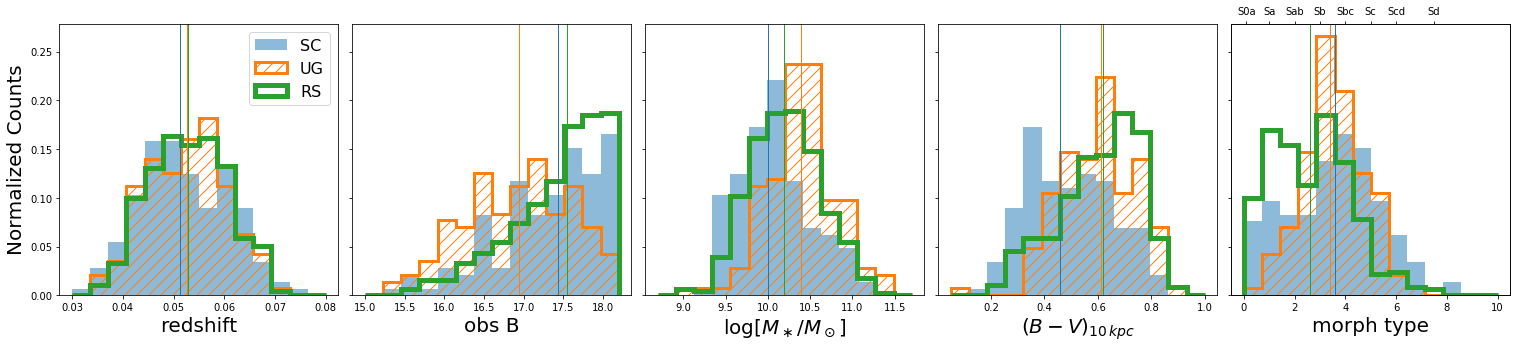}
    \caption{Normalized distributions of redshift, observed B magnitude, stellar mass, rest-frame (B-V) color, and morphological type for the ram pressure stripped candidates (SC, light blue), the unwinding galaxies (UG, orange) and the reference sample (RS, green) samples. Vertical lines represent median values. 
    \label{fig:cfr_distributions2}}
\end{figure*}

\subsection{The galaxy properties } \label{sec:properties}

\begin{figure*}
    \centering
    \includegraphics[scale=0.43]{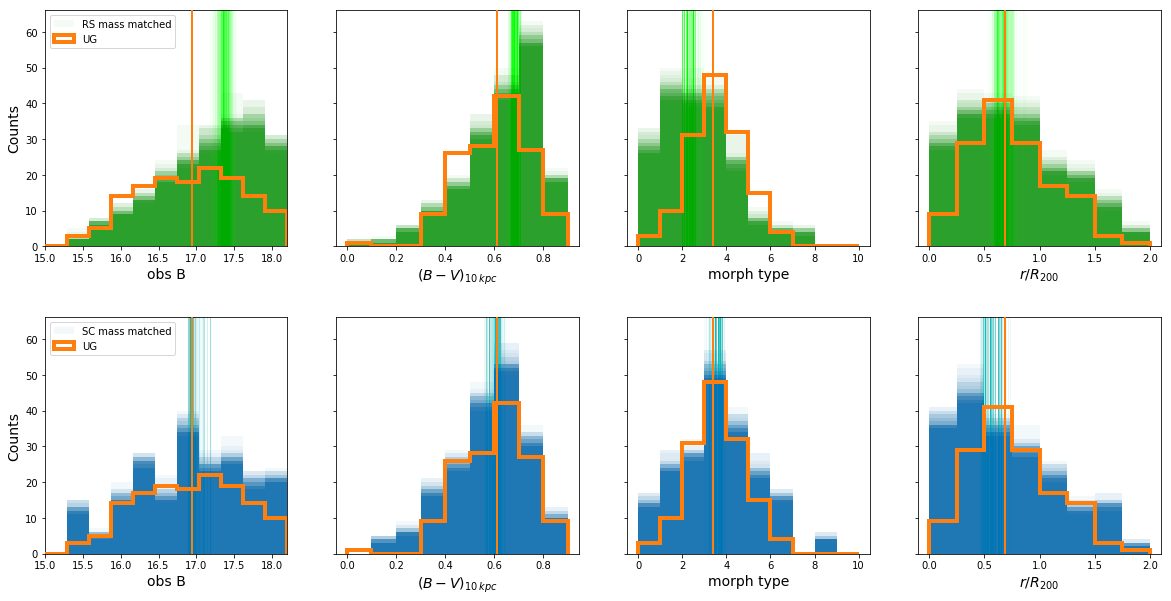}
    \caption{Distributions of observed B magnitude, rest-frame (B-V) color, morphological type and clustercentric distance in units of R$_{200}$ along with median values for the UG sample (orange) and RS (top, green), and SC (bottom, blue) once their mass distribution has been matched to the one observed for the UG. All the 1000 random extractions are reported, along with their median value.
    \label{fig:cfr_mm}}
\end{figure*}

\begin{figure}
    \centering
    \includegraphics[scale=0.6]{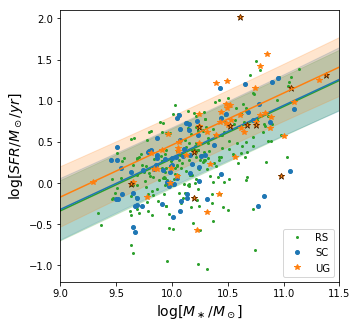}
    \caption{Star Formation Rate - Mass relation for galaxies in the different samples. Only the subset with available SFR values are considered (see text for details). Colors and symbols are as in Fig.\ref{fig:pps}. Solid lines and filled areas show the linear best fit to the relation, along with the 1$\sigma$ error. For all the samples, the slope has been fixed to the slope of the full sample (RS+UG+SC).
    \label{fig:sfrmass}}
\end{figure}

We now compare the galaxy properties of the different samples in Fig.\ref{fig:cfr_distributions2}. SC, UG and RS span a very similar redshift range, and a KS test is not able to detect any difference between them. Focusing on the B band magnitude, UG are systematically brighter (by $\sim 0.5$mag) than the other populations, indicating that either this population is indeed distinct, or simply that our ability to identify this class of object is limited to bright galaxies. In addition, detected UG are typically face on, so  they can be brighter because in galaxies with some inclination one side of the galaxies is obscured and therefore galaxies might appear fainter. 
A KS test retrieves differences between all populations, at more than 1\% level. Also stellar mass distributions are clearly different, with SC being the least massive population (KS p$_{value}$=0.02 when comparing SC to both UG and RS) and UG the most massive one (KS p$_{value}$=6$\times10^{-7}$ when comparing UG and RS).  Still, UG cover a wide range of galaxy masses, from $10^9 \, M_{\odot}$ to $10^{11.5} \, M_{\odot}$.  SC differentiate from the other two populations also in terms of colors: they are significantly bluer than UG (KS p$_{value}$=1$\times10^{-6}$) and RS (KS p$_{value}$=2$\times10^{-7}$), which instead are indistinguishable, both according to the KS test and given their median value. Finally, the three populations are also different in terms of morphological distributions and the KS test supports this finding with high confidence level (KS p$_{value}\sim$1$\times10^{-10}$ when comparing RS to both UG and SC, KS p$_{value}$=0.02 when comparing UG to SC). RS galaxies are dominated by late-type galaxies of earlier morphology: in clusters there are very few undisturbed galaxies of later morphology. In contrast, the distribution of UG is much narrower and peaked around Sb-Sbc galaxies (morph type 3-4). This finding is in line with expectations: a galaxy must have a clear spiral morphology so that its arms can be unwound.  SC are characterized by a rather broad distribution and skewed towards higher values of morphological type: galaxies undergoing stripping must have a large of gas reservoir that can be sensitive to the ram pressure.  

Results remain qualitatively the same both when we consider only galaxies with no disturbances ($F_d$=0) and also when consider only the most secure and clear candidates (JClass and UClass 3,4,5, respectively).

The results discussed above can be driven by the different mass distribution characterizing the different samples. To remove this effect, in Fig.\ref{fig:cfr_mm} we compare galaxy properties when the different mass distribution is taken into account. We take as reference the UG sample and extract from the SC and RS samples, respectively, 1000 subsamples characterized by the same mass distribution as the UG sample and plot the observed magnitude, color and morphological type distribution of these realizations. We also plot median values of all the realizations, that give a sense of the spread of the distributions. All differences between UG and SC disappear, as median values of the quantities are very similar and KS test runs are not able to detect differences when comparing the observed UG quantity and the mass-matched SC one. In contrast, differences between UG and RS remain: median values are systematically different and the KS test is able to detect differences between the distributions in 99\% of the realizations. Controlling for stellar mass,  the UG sample is still brighter, bluer and have a later morphology than the RS.

The right panels of Fig.\ref{fig:cfr_mm} compare the radial distribution of galaxies in the different subsamples, for subsets of galaxies with the same mass distribution. In this case, UG and RS galaxies have indistinguishable distributions, while the differences seen in Fig.\ref{fig:pps} between UG and SC become even more evident: UG are systematically found at much larger distances than SC 
\citep[see, e.g., Fig. 11 and 12 in ][]{Bellhouse2021}. Performing a KS test on each of the 1000 realizations, we find that distributions are extracted from different parent samples with high significance in 70\% of the cases. 

Stellar mass therefore seems not to be the only parameter driving the appearance of UG, but it is the main source of differences between UG and SC appearances. Still, minimizing the role of the stellar mass, we find UG are still found at larger clustercentric distances, compatible with the assumption that UG are an early stage of SC.  This is in agreement with what found by \cite{Gullieuszik2020} who showed that the stellar mass is not the only parameter driving the amount of gas that is stripped by ram-pressure.

Finally, we investigate how the three samples distribute on the SFR-mass relation.
The galaxy current SFR was derived applying our spectrophotometric tool SINOPSIS to the available optical spectroscopy, as described in \citep{Fritz2007, Fritz2017}. The model performs a non-parametric full spectral fitting of the continuum shape and of the main emission and absorption lines, deriving a star formation history. The ongoing SFR is constrained from the fluxes of the emission lines and the blue part of the spectrum, and dust extinction is taken into account. Being obtained from multifiber spectroscopy, the SFR estimate refers to the central region of galaxies that is covered by the fiber (that has a diameter of 2.1 arcsec, covering the central 1.7–2.8 kpc at the WINGS redshifts), and is then extrapolated to a total SFR value assuming the same mass-to-light ratio within and outside of the fiber. The mean correction factor is 6, with a standard deviation of 4.8 (\citetalias{Poggianti2016JELLYFISHREDSHIFT}). We do not have SFR estimates for those galaxies whose redshift comes from the literature, but we can assume that the subsample for which this quantity is available is just a random subset of the whole sample, so results should be unbiased. Overall, a SFR measurement is available for 317 (=69\% of the full sample) RS, 85 (60\%) UG and 90  (=62\%) SC.
To determine the fit, we obtain the slope of the relation fitting all the galaxies together, and then determine the intercept for each population separately. 
\citetalias{Poggianti2016JELLYFISHREDSHIFT} already showed how SC tend to be located above the best fit to the relation traced by all cluster members. Figure \ref{fig:sfrmass} shows that we do not recover significant differences between SC and RS. This can be due to the different sample selections:  \citetalias{Poggianti2016JELLYFISHREDSHIFT} compared their candidates to the full cluster population, including also redder late- and early-type galaxies,  while here we focus only on blue late-type cluster members with no clear signs of mergers, interactions, etc. In contrast, we find that UG do show a tentative excess in SFR at any give stellar mass with respect to the RS. Their relation is $\sim 0.15$ dex above that of the RS sample. 
UG with  the strongest  unwinding features (UClass 3,4,5) do not seem to lie in specific regions of the plane. We reach similar conclusions both letting the slope vary among the different samples and computing the running median instead of the linear fit.

Note that AGN have not been identified in the OMEGAWINGS spectra, therefore AGN can be present in the sample and the measured SFR might be overestimated in these galaxies. We visually inspected the spectra of UG located 1$\sigma$ above the fit of the relation and found that only the galaxy with the highest SFR has a spectrum typical of Seyfert 1, with very broadened emission lines.  Removing the galaxy does not change the presented results. A detailed analysis of the possible correlation between unwinding features and presence of AGN is beyond the scope of this work. 

\section{The incidence of ram pressure stripping candidates in local clusters} \label{sec:results}
Having characterized the properties of UG with respect to the other populations, we are now in the position of drawing general conclusions on the incidence of galaxies most likely affected by ram pressure stripping in clusters. 

We remind the reader that our samples are based on visual inspection of optical images and only a subset of the galaxies used in this paper have been indeed confirmed to be ram pressure stripped through integral field spectroscopy. In addition, we still have not quantified  how often unwinding features in clusters are produced solely by ram pressure stripping. Nonetheless, if many UG are suffering ram pressure, this exercise gives us some idea on how much they contribute to the total ram pressure stripping fraction in local clusters and we will use the different samples to provide some confidence levels. 

In our sample, the fraction of SC over the total  blue, non interacting, late-type galaxies is 0.19$\pm0.01$, with quoted uncertainties measured as binomial errors. Considering only galaxies with no disturbances ($F_d=0$), this fraction rises to 0.24$\pm 0.02$. Similarly, the fraction of UG is 0.19$\pm0.01$, it raises to 0.28$\pm0.02$ when only $F_d=0$ galaxies are taken into account. If we assume that both SC are UG are {\it all} produced by ram pressure stripping, their incidence over the total population is  0.38$\pm0.02$.
The same fraction becomes 0.52$\pm0.02$ when considering galaxies with  $F_d=0$, meaning that more than half of the blue late-type cluster population would be affected by ram pressure stripping in a way that is noticeable from optical imaging.

As seen in the previous section, the different populations are characterized by different galaxy properties and location within the clusters. In the next section we will therefore investigate how these fractions depend on some quantities. 

\subsection{Dependence on galaxy properties}

\begin{figure*}
     \centering
     \includegraphics[scale=0.35]{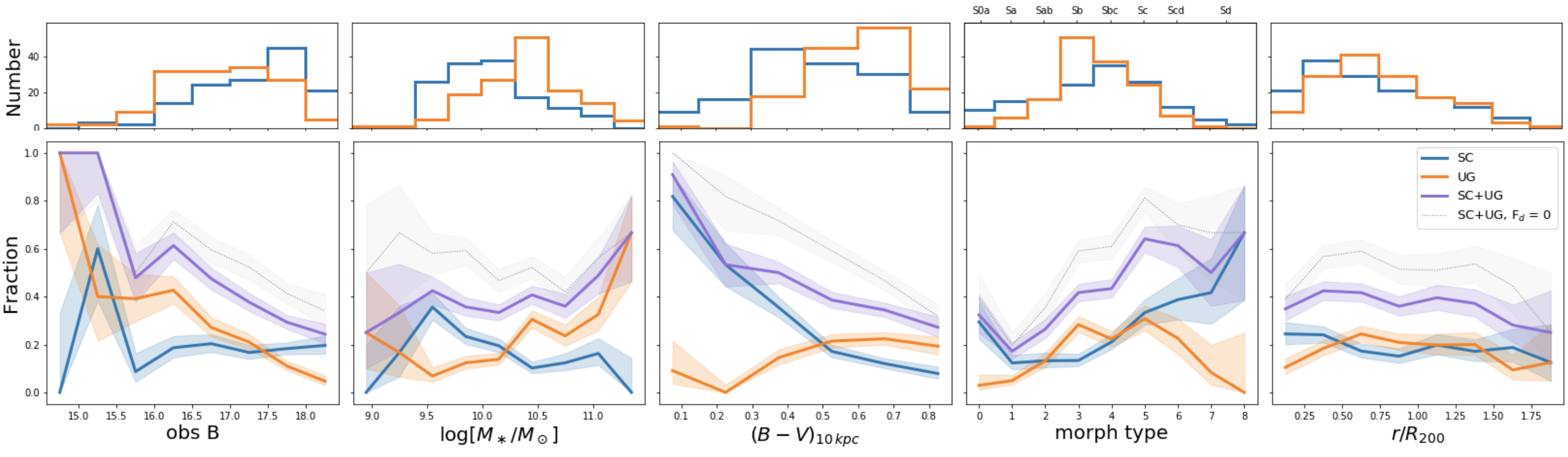}
     \caption{Fractions of galaxies as a function of galaxy properties. From left to right: observed magnitude, stellar mass, B-V color, morphological type, clustercentric distance. Light blue lines refer to the SC sample, orange line to the UG sample, purple lines to the sum of UG+SC. The light dotted black line is obtained by considering only galaxies with $F_d$=0. Errors are binomial. The upper inset show the number of galaxies per bin. 
     \label{fig:fractions}}
 \end{figure*}

Figure \ref{fig:fractions} shows the incidence of the two populations with respect to the entire cluster population of non-interacting blue, late-type galaxies (SC+UG+RS) as a function of observed magnitude B, stellar mass, $(B-V)_{10 \, kpc}$ color, morphological type, clustercentric distance. Under the assumption the both populations are produced by ram pressure stripping, we also investigate trends for the sum of the two. The vast majority of bright galaxies show signs of stripping: for B$<$16 more than 50\% of them are candidates for being stripped. UG dominate this fraction: they are 100\% for B$<$15.5. Their incidence decreases with magnitude and for B$>17.5$ the UG fraction is negligible. In contrast, the relative importance of SC is maximum at B$\sim$15, then it drops and stays constant ($\sim 20\%$) at all magnitudes. Note that in the first bin there are only few galaxies, therefore results must be taken with caution. 
The incidence of SC+UG increases with mass and ranges from 20\% at $\log M_\ast/M_\sun =9$ to 60\% at $\log M_\ast/M_\sun =11$. This trend is mainly driven by UG: the trend for SC is instead the opposite, with their fraction decreasing from  $\log M_\ast/M_\sun =9.5$ (where they represent $\sim 40\%$) to $\log M_\ast/M_\sun >11$, where they are almost absent. The incidence of UG is negligible for $(B-V)_{10 \, kpc}<$0.3 and then its almost independent on galaxy color, while that of SC strongly depends on color: for $(B-V)_{10 \, kpc}<$0.3 SC represent more than half of the cluster population while their fraction drops toward redder colors.
Morphology also drives the incidence of galaxies possibly undergoing ram pressure stripping: the incidence of SC+UG monotonically increases going from the earlier to the later types of spirals: galaxies with a morphology later than Sc have a 60\% chance of being stripped. UG dominate the fractions for intermediate types, while SC are more frequent at earlier and later types. 
Finally, only mild trends are evident with clustercentric distances: the incidence of both UG and SC slightly decreases with increasing distance. 

Considering only galaxies with $F_d=0$ (grey line in Fig. \ref{fig:fractions} only for SC+UG), trends stay the same, but fractions increase by $\sim 20\%$ in each bin. This is due to the fact that overall most of the SC and UG galaxies have $F_d=0$, meaning that we label them as ram pressure stripping only candidates and we do not detect other concurring mechanisms. In contrast, many of the RS galaxies are not really undisturbed, but might be affected by  non -identified processes. In clusters, the incidence of truly undisturbed blue, late-type galaxies is very low. We remind the reader that galaxies in the RS can be either first infallers (most likely not disturbed yet) or  galaxies that very recently quenched and had no time to change color yet, but that could have undergone some other mechanisms or they could also be on orbits that never expose them to ram pressure stripping. 

To summarize, assuming that both UG and SC are the result of ongoing ram pressure stripping, we can conclude that bright, massive and blue late spiral galaxies have an extremely high chance of showing evidence of stripping: indeed, 60\% of galaxies more massive than $\log M_\ast/M_\sun =11$ show signatures compatible with ram pressure stripping, as also galaxies bluer than $(B-V)_{10 \, kpc}<0.2$. Assuming instead that only a negligible part of the UG are indeed the by-product of ram pressure stripping, we find that galaxies with  $\log M_\ast/M_\sun =9.5$ have a higher chance of being stripped: 40\% of the cluster galaxies at these masses are SC. Morphology is still an important parameter to take into account: almost half of galaxies with a very late type morphology show signs of stripping.

\section{Discussion} \label{sec:disc}
The main goal of this paper is to quantify for the first time the incidence of galaxies showing signs of ram pressure stripping in local clusters from optical imaging. 

Based on GASP data, Poggianti et al. (in prep) have shown that the "success rate" of an optical imaging search of ram pressure stripped galaxies in local clusters is $\sim 85\%$. This means that about 85\% of the optical imaging stripping candidates are confirmed to be ram pressure stripped galaxies based on integral-field spectroscopy revealing the stripped ionized gas. In what follows we therefore take this percentage into account when estimating the incidence of SC, whose fraction becomes 0.16$\pm0.01$.

To have a complete picture, we have also performed a systematic search of galaxies showing unwinding features in local massive clusters, under the assumption that these can in some cases be produced by ram pressure stripping, complementing previous searches of galaxies with unilateral stripped debries. 
This assumption is supported both by observational studies \citep{Bellhouse2017, Bellhouse2019, Bellhouse2021} and by theoretical expectations \citep{SchulzStruck2001, Roediger2014, Steinhauser2016}, but previous visual identification efforts aimed at selecting ram pressure stripped candidates did not take this category of galaxies in consideration (\citetalias{Poggianti2016JELLYFISHREDSHIFT}, \citealt{Roberts2021a, Durret2021}).

However, unwinding features could also be related to tidal interactions \citep{Pettitt2016, Pettitt2017}. Simulations have shown that in this case, the features are expected to be fairly symmetrical for  high speed encounters, rather asymmetrical in case of low speed encounters (like  M51) \citep{Byrd1990}. In the latter case, the galaxy appearance could be similar to ram pressure stripped cases, but the nearby companion should be still visible, allowing us to distinguish between the two mechanisms \citep{Smith2021}.

Further,  unwinding features could also be related to galaxy internal properties, such as bulge mass and gas density (\citealt{Davis2015, Davis2017}, but see \citealt{Hart2017, Hart2018, Masters2019}).
Also galaxy bars could play a role: \cite{Masters2019} found that the presence of a strong bar tends to correspond to more loosely wound arms. It is still not clear though if the presence of a strong bar induces the unwinding, or if also bars can form during a ram pressure stripping event that destabilizes the gas and helps simultaneously the development of a bar and unwinding arms. In the OMEGAWINGS sample, the characterization of the bars is currently underway, therefore a joint analysis between unwinding features and bars is deferred to a future work.

Given the unclear origin of the unwinding features, we have considered two different extreme scenarios: on one side galaxies showing unwinding features are all undergoing stripping, on the other hand the contribution of UG to the total stripping population is negligible, therefore only galaxies with clear tails are representative of galaxies undergoing ram pressure stripping. Depending on the adopted scenario, the incidence of ram pressure stripped candidates  over the cluster population of non -interacting blue late-type galaxies  ranges from  35 to 15\%. These two cases  should bracket the real frequency of galaxies showing signs of ram pressure stripping at any given time in optical bands. 

We remind the reader that in our analysis we have excluded clear cases of mergers and strong interactions, therefore all the quoted fractions refer to the cluster population of non-interacting objects. If we instead include also these galaxies to obtain the fraction of the global cluster population of blue late type galaxies, we obtain that  the fraction of SC is  0.14$\pm0.01$, the fraction of UG is  0.15$\pm0.01$ and their combined fraction is 0.29$\pm0.02$.

If all UG are due to ram pressure stripping, it would be interesting to understand if the unwinding appearance can develop only in an initial stage of the stripping phase and if some galaxies will never develop proper tails, being the unwinding features the only visible sign of the stripping. Only targeted simulations will be able to answer this question. It would also be important to understand if there is a main parameter driving the appearance. This could be either the properties of the orbits on which the galaxy is located, the properties of the hosting cluster (its mass, relaxation state, conditions of the ICM) or galaxy mass. In a very massive galaxy, the anchoring force could be sufficiently high to contrast the strength of the ram pressure and prevent the formation of galaxy tails \citep[but see][]{Poggianti2019}, while low mass galaxies more easily develop tails \citep[e.g.,][]{Grishin2021}.   All these considerations are similar to those faced by \cite{Gullieuszik2020} to explain the development of tails in SC. They indeed found that the mass of the galaxy, its position and velocity in the host cluster, and all of the parameters defining the distribution of the ICM density are all to be considered to properly account for the development of tails, more specifically the amount of star formation in the tail \citep[see also][]{Jaffe2018}. 

Ours is among the first attempts to quantify the importance of ram pressure stripping in cluster evolution. 
In simulations, \cite{Yun2019} have quantified in the TNG100  the number of $z<0.6$ galaxies with $M_\ast>10^{9.5} M_\odot$ showing signs of ram pressure stripping in clusters (host halo masses in the range $10^{13} \leq M_{200c}/M_\odot \leq 10^{14.6}$).
They performed a visual selection, along the line of the one performed in \citetalias{Poggianti2016JELLYFISHREDSHIFT}, and found that 31\% of cluster satellites show gaseous tails stemming from their main luminous bodies. In the same mass and clustercentric distance range ($r/r_{200}<1$) we roughly obtain a similar fraction of SC: 0.38$\pm0.02$. However, it is worth stressing that 1) the galaxy identification is based on different criteria, as they inspect  color-maps of the gas column density, rather than optical images; 2) the clustercentric distribution they  obtain from simulations does not match the one presented in Fig. \ref{fig:pps}, they indeed fail to detect ram pressure stripped galaxies close to the cluster center ($r/r_{200}<0.2$), where instead the most beautiful jellyfish galaxies are found in observations  \citep[e.g.,][]{Gullieuszik2020, Moretti2021}. They interpret their result with the fact that possibly, the stripping mechanisms have already depleted the gas from the galaxies that are found at the time of observation at small clustercentric distances. 

Observationally, results are strictly linked to the image depth and wavelength considered, as different selections highlight different ram pressure stripping signatures and therefore can (dis-)favour ram pressure stripped galaxies at different phases.
Only works simultaneously considering many different wavelengths might give a comprehensive view of the stripping, but large samples of galaxies with complementary observations from the X, UV, sub-mm, HI to the radio are  not available yet.

For example, radio continuum observations are sensitive to synchrotron emission from cosmic rays accelerated by supernovae. For galaxies experiencing strong ram pressure, these cosmic rays can be stripped out of the galaxy and detected as ram pressure stripped tails in the radio continuum \citep[e.g.,][]{Gavazzi1987, Murphy2009, Chen2020}, giving reliable identifications of jellyfish galaxies. \cite{Roberts2021a}
used the LOFAR Two-metre Sky Survey (LoTSS) to search for jellyfish galaxies at 120-168 MHz radio wavelengths. They find that the frequency of ram pressure stripped  galaxies among star-forming galaxies (sSFRs $>10^{-11} yr^{-1}$) is relatively constant from cluster to cluster and that  ram pressure stripped  galaxies are preferentially found at small clustercentric radius and large velocity offsets within their host clusters. \cite{Roberts2021b} extended the previous analysis considering SDSS groups and clusters ranging in mass from $\sim10^{13} - 10^{14.5} M_\odot$. They find that ram pressure stripped galaxies are most commonly found in clusters, with the frequency with respect to star forming galaxies decreasing towards the lowest-mass groups, with fractions ranging between 5 to 20\% depending on environment. 

Optical selection instead favours galaxies with star formation in the tail.  This selection might underestimate the total number of galaxies, as it has been shown that some galaxies at peak stripping do not have star formation in the tail \citep[e.g.,][]{Boselli2016, Laudari2021}. In addition, optical wavelengths allow us to detect only the galaxies at an advanced stage of stripping, as new stars have to have time to form out of the stripped gas \citep[e.g.,][]{Poggianti2019,Gullieuszik2020}. Nonetheless, very advanced stages of stripping, visible as truncated disks \citep[e.g.,][]{Fritz2017}, are also disfavoured by our selection criteria. 

All these biases have  to be taken into consideration when interpreting the results  presented in our work. 
Optical imaging  has been exploited also by \cite{Roberts2022}, who  studied a sample of ram pressure candidates, visually identified from the Canada-France Imaging Survey, in tenths of SDSS groups and clusters ranging in mass from $\sim10^{13} - 10^{14.5} M_\odot$. They recovered similar fractions to those by \cite{Roberts2021b} and they also find that the frequency of ram pressure candidates is highest for low- mass galaxies (5\% of the sample). 
The selection criteria adopted by \cite{Roberts2022} are significantly different from ours, so the parameter space (morphology, color, magnitude) covered by their and our study is significantly different, and results can not be directly compared, though overall findings are in agreement. 

To really quantify the importance of the phase we observe at optical wavelengths and obtain a general census of galaxies undergoing ram pressure stripping in local clusters  we need to take into account the duration of the visibility of this phase in galaxies. 

According to simulations, blue galaxies infalling from the field reach pericenter in about 2.4 Gyr (\citealt{McGee2009}, see also \citealt{Oman2013})
and we can hypothesize that by the time they get there they lose most of their gas. In the first part of this time frame, they can start feeling the ram pressure on act and their neutral gas content, less bound, starts being stripped. No signs at optical wavelengths are visible yet, but the stripping has started \citep{Tonnesen2010}. In some galaxies ram pressure stripping might be able to unwind the spiral arms \citep{Bellhouse2021}, in others it just strips the gas that at some point collapses and starts forming new stars \citep[e.g.,][]{Poggianti2019}. In this phase, tails shine at optical wavelengths, as OB massive stars are very bright. We can assume the visibility of this phase lasts of the order of $\sim 6\times$ 10$^8$ yr \citep{Fumagalli2011, Poggianti2019}. Stripping and star formation consume the available gas and galaxies appear as truncated disks first \citep{KoopmannKenney2004a, Fritz2017} and then become passive, showing k+a spectra \citep{Vulcani2020}. Galaxies typically maintain blue colors for 0.5 Gyr after becoming passive and then move to the red sequence \citep{poggianti04}.  
In these phases tails are not visible anymore at optical wavelengths. 
So to summarize, an infalling galaxy  maintains its blue color for at least 2.4+ 0.5 $\sim 3$ Gyr since infall and a tail is visible only for 6$\times$ 10$^8$yr. We can assume that the sum of RS+SC+UG samples corresponds to all the non-interacting blue late-type galaxies in the clusters and that the tailed galaxies are either only SC or SC+UG. In the first case, $\sim$15\% 
of the blue cluster galaxies currently show signs of stripping, but since the visibility phase is $6 \times 10^8/3\times 10^9 \sim $0.2, the total amount of galaxies undergoing stripping is  15\%
/0.2 $\sim$ 75\%.  In the second case,  the blue cluster galaxies currently show signs of stripping are even more, on one side suggesting  that all blue cluster galaxies undergo a stripping phase during their life in clusters and on the other side indicating that most likely we are overestimating the number of ram pressure stripped galaxies only using optical imaging. 

This rough calculation shows that ram pressure stripping is an extremely important process able to affect the vast majority of cluster galaxies. We remind the reader again that not all ram pressure stripped galaxies can show tails visible at optical wavelengths.

\section{Conclusions} \label{sec:conc}
In this paper we have provided the first census of galaxies undergoing ram pressure stripping in local cluster galaxies as obtained from optical imaging, 
starting from the catalog of ram pressure stripping candidates released by  \citetalias{Poggianti2016JELLYFISHREDSHIFT}. In order to obtain a comprehensive view on this phenomenon, we have first carried out a systematic visual search of galaxies showing unwinding arms in the WINGS and OMEGAWINGS clusters. This effort was dictated by the recent findings by \cite{Bellhouse2017, Bellhouse2019, Bellhouse2021} that point to ram pressure stripping as responsible of unwinding features in at least some cluster galaxies, contrary to usual expectations that only tidal features can produce unwinding arms. Unwinding galaxies with no clear tails are though absent by construction in  \citetalias{Poggianti2016JELLYFISHREDSHIFT}.

We restricted our search to cluster members within 2$R_{200}$, with B$<18.2$ and blue colors.  Overall, we identified 143 galaxies (UG) with a different degree of unwinding evidence, indicated in the UClass. UClass 5 galaxies have the highest evidence of unwinding, UClass 1 the lowest. In the sample we identified  2 UClass 5, 7 UClass 4, 26 Uclass 3, 56 Uclass 2 and 52 UClass 1. Of these, 22 have the disturbance flag set to 1. We release the catalogs and images of these galaxies.

During the visual inspection, we have also identified 35 ram pressure stripping candidates that were missed by \citetalias{Poggianti2016JELLYFISHREDSHIFT}.
We here release the properties of these galaxies. Combining this sample to the  \citetalias{Poggianti2016JELLYFISHREDSHIFT} one above the same magnitude, color, morphology limits, the stripping candidate (SC) sample includes 145 galaxies.  We also assembled a reference sample (RS), made of blue late-type undisturbed cluster galaxies, for a total of 459 galaxies. We on purpose excluded clear cases of mergers and strong interactions from this sample, as these would hinder a clear detection of possible signs of ram pressure stripping. 

We have then characterized the sample of unwinding galaxies, both in terms of frequency and location within the clusters and of their properties and compared it to the other samples. The main results can be summarized as follows.
    \begin{itemize}
        \item Clusters host a varying number of UG, from 0 to 11. Considering the number of blue, late-type members in each cluster, the incidence of UG ranges from 0 to 80\%, with an average value of 20\%. The number and fraction of UG seem not to depend on cluster properties, such as $L_X$ and $\sigma_{cl}$.
        \item UG are found at all clustercentric distances and are characterized by a wide range of velocities. Compared to SC, they are located at slightly larger distances, compatible with the expectation that they fell into cluster more recently than SC.
        \item UG are typically brighter, redder, more massive than SC and have a Sb-c morphology, while SC have a broader morphology distribution. UG have a similar $(B-V)_{10 \, kpc}$  color of RS, but differ in all the other properties. When controlling for stellar mass, differences with SC disappear, while those with RS stay significant, suggesting stellar mass plays some role in determining whether a galaxy can develop any stripping signatures or not. 
        UG and SC in this respect are consistent with being a unique population, and stellar mass is the main parameter determining unwinding arms or tails in ram pressure stripped galaxies.
        \item Focusing on the star forming properties, UG are preferentially located above the relation obtained by RS galaxies. Assuming the two populations can be described by the same slope of the fit, the intercept of UG is 0.15 dex above that of the RS.
    \end{itemize}
 
Having identified and characterized the three samples, we have estimated the incidence of UG and SC over the cluster population. They each represent $\sim 15-20\%$ of the total population of non-interacting blue late type cluster galaxies. Assuming that they both are undergoing ram pressure stripping we can conclude that at any given time about 35\% of the non-interacting blue late type cluster population show signs of stripping in their morphology at optical wavelengths.
 
The incidence of these populations depend on galaxy properties. Considering only galaxies with $\log M_\ast/M_\sun >11$, the fraction of galaxies with signs of stripping 
becomes 60\%. Similarly, also color and morphological type play a role. In contrast, the fraction of these galaxies is independent on clustercentric distance.

Making rough assumptions on the duration of the tail visibility and of the time needed for cluster galaxies to become red, we infer that 
potentially all cluster galaxies undergo a stripping phase visible in optical imaging during their life. Ram pressure stripping is therefore  a crucial mechanism in cluster galaxy evolution. 

\section*{Acknowledgements}
We thank Daniela Bettoni, Stephanie Tonnesen, Sean McGee and the rest of the GASP team for useful discussions.  Based on observations collected at the European Organization for Astronomical Research in the Southern Hemisphere under ESO programme 196.B-0578. This project has received funding from the European Research Council (ERC) under the European Union's Horizon 2020 research and innovation programme (grant agreement No. 833824).  We acknowledge financial contribution from the grant PRIN MIUR 2017 n.20173ML3WW\_001 (PI Cimatti), from the INAF main-stream funding programme (PI Vulcani) and from the agreement ASI-INAF n.2017-14- H.0 (PI A. Moretti). Y.J. acknowledges financial support from ANID BASAL project No. FB210003 and FONDECYT Iniciaci\'on 2018 No. 11180558. J.F. acknowledges financial support from the UNAM- DGAPA-PAPIIT IN111620 grant, México.


\bibliography{references}{}
\bibliographystyle{aasjournal}



\end{document}